\documentclass{IEEEcsmag}

\usepackage[colorlinks,urlcolor=blue,linkcolor=blue,citecolor=blue]{hyperref}

\usepackage{upmath}
\usepackage{hyperref}
\usepackage{graphicx}
\usepackage{multirow}
\usepackage{color}
\usepackage{xcolor}
\usepackage{soul} 
\usepackage[normalem]{ulem}

\jvol{XX}
\jnum{XX}
\paper{8}
\jmonth{Feb/Mar}
\jname{IT Professional}
\pubyear{2023}

\begin{document}

\editor{}
\title{A blockchain-based e-goverment service for Quantity Surveyors}

\author{Ángel F. Alcaide}
\affil{COAAT's Computer Technician, Albacete (Spain)}
\author{Carlos Núnez-Gómez}
\affil{High-Performance Networks and Architectures Group (RAAP), Albacete Research Institute of Informatics (I3A), University of Castilla-La Mancha}
\author{Francisco M. Delicado, Carmen Carrión, M. Blanca Caminero}
\affil{Department of Computing Systems, University of Castilla-La Mancha}

\markboth{Department Head}{Paper title}


\begin{abstract}
In Spain, quantity surveyors are entitled to carry out official cadastral surveys, attestations, and certificate issuing according to a well-defined professional code. Official Associations of Quantity Surveyors and Technical Architects (COAAT) are responsible for endorsing the documentation related to actions performed on buildings. An e-platform that enables immutability, traceability, and a unique property record among all the Spanish COAATs, with an affordable cost, is essential to streamline the involved processes. 

The blockchain technology and smart contracts have recently emerged as promising solutions for e-government services due to the inherent features provided by the technology. In this paper, we identify the design goals and propose a blockchain-based e-government system for the electronic management of the documentation generated, submitted, and validated by the Spanish COAATs, namely, the COAATChain. The proposal has been deployed and evaluated on the Binance testnet blockchain, in order to assess its affordability.

\end{abstract}

\maketitle
\chapterinitial{All over the world,} governments and government agencies are now implementing projects and initiatives for digital transformation and to provide effective digital services or e-government~\cite{ebsi, estonia}. 
In general, e-government services improve citizen experience and save costs. 
Users expect high-availability and fault-tolerance from the e-government services which increase their satisfaction. 
Nowadays, emerging technologies and innovative practices help governments to meet these requirements.
In particular, blockchain-based smart contracts are one of the most up-and-coming solutions with significant features for e-government services~\cite{HEWA2021}.

In this paper, we will focus on the Spanish Official Association of Quantity Surveyors and Technical Architects (COAAT, by its Spanish acronym) which is in charge of endorsing the documentation of the projects, referring to new building works, refurbishments or energy efficiency in order to grant them an additional validity for their subsequent presentation to public bodies, administrations, city councils or community boards. Surveyors carry a lot of responsibility according to a well-defined professional code. For example, they carry out official cadastral surveys and may perform attestations and issue certificates. Some types of projects, such as building demolitions, require mandatory approval by a COAAT.

For many years it was compulsory to present all the physical documentation in person at the offices of the professional associations, leave it for a few days for review and validation if appropriate, then collect the approved documentation and finally present it to the corresponding public administration. It should be noted that each professional association represents a demarcation, which coincides with the provincial administrative division. There are several professional associations within a country.
In recent times, both professional associations and public administrations have made technological efforts to digitise this process. For example, the digital certificate system is now supported, which saves time and travel.
However, there are still some important challenges and requirements for improvement. 

A digital service for the Spanish COAAT should tackle the following outstanding features:

\begin{itemize}
\item Immutability: Prevent manipulation of property records by a central authority or a third party.
\item Common and unique property record: Avoid double property records and share information among different official associations.
\item Record traceability, timestamp, and provenance: Keep track of property records.
\item Affordability: Automate and streamline the access, processing, and validation of property records at reduced cost.
\item Resiliency: Provide a system setup with high availability and fault-tolerance.
\end{itemize}

Taking into account the current-state analysis, our main objective is to design a decentralized e-government system for the Spanish COAAT that fulfills the key features mentioned above.
In a nutshell, our proposal uses the following solutions:

\begin{itemize}
    \item The design of an e-governance system, called COAATChain, based on role definitions. 
    \item The implementation of the system with smart contracts.
 
 \end{itemize}

A methodology similar to the Trustable DApp Modelling (T-DM) methodology proposed in~\cite{udokwu2021} has been followed to derive the system implementation from the system value proposition. As a result, the COAATChain system has been designed, deployed, and tested on a blockchain-based platform to meet the initial requirements.

The rest of the paper is organized as follows. First, we present the related work. We then introduce COAATChain stakeholders and its first level goals (FLGs), and also describe the most important implementation details. Then, a design evaluation of the approach is discussed. And finally, conclusions and future work are presented.

\section{Related work}

Smart contracts are digital programs whose code is stored on the blockchain~\cite{ferdous2020blockchain}. These programs are applicable in different application domains (i.e. financial applications, e-government, 
insurance, auditing, or healthcare). 

Within the e-government domain, we find blockchain-based solutions for electronic voting which ensure authentication, anonymity, accuracy, and verifiability~\cite{garg2019comparitive}, or distributed ledger based approaches to automate the intellectual property licensing payments~\cite{Tietze2020}. 
We also find modeling for the semi-automated translation of human-readable contract representations into computational equivalents that enable the encoding of laws~\cite{frantz2016institutions}.
In~\cite{chiang2018exploring}, the authors present ChainGov as a collaborative blockchain and smart contract integrated platform to build trust between immigrants, governments, and other institutions informing about the flow of money as well as real-time visibility of all transactions. 
Also, blockchain has been proposed as the core to support global vaccination strategies for COVID19, harnessing its decentralization and global unique identification~\cite{Costa2022}.
One of the countries that have advanced the most in terms of e-government is Estonia~\cite{estonia}. They have created the concept of digital country, where many procedures can be carried out by electronic means. In particular, an ad-hoc scalable blockchain was developed to ensure the integrity of data stored in government repositories and to protect its data against insider threats~\cite{estoniaPWC,Semenzin2022}.

Moreover, the European Commision is building its own blockchain infrastructure, the European Blockchain Services Infrastructure (EBSI), \emph{``to leverage blockchain to create cross-border services for public administrations, businesses, citizens, and their ecosystems to verify the information and make services trustworthy''}~\cite{ebsi}. EBSI focuses on providing verifiable credential services, and traceability of documents, among others. There are several ongoing pilot projects to demonstrate its usefulness and increase awareness and adoption of the technology~\cite{Lykidis2021}.

\section{COAATChain stakeholders and system value proposition}\label{sec:svp}

Based on expert knowledge, different stakeholders (i.e., user roles) have been identified for the proposed COAATChain system:

\begin{itemize}
    \item \textbf{Role~0 (System administrator)}: Responsible of the initial deployment of the system and the registration of the COAATs.
    
    \item \textbf{Role~1 (COAAT administrators)}: They can access, validate or reject dossiers, register new properties, and add quantity surveyors (Role~2) and other agents (Role~3) into the system. 
    
    \item \textbf{Role~2 (COAAT staff)}: They include quantity surveyors and technical architects, that can register a new property, create a new construction dossier associated with an existing property, and include documentation to the dossiers previously created by them. COAAT staff cannot modify the dossiers created by other surveyors, even though he/she can see that they exist.
    
    \item \textbf{Role~3 (Read-only users)}: 
    Notaries, court members, public administration workers, and any legal entity that by law has the right to access the information in the construction dossiers stored in the system. 
\end{itemize}

\begin{figure*}[ht!]
    \centering
    \includegraphics[width=1.0\textwidth]{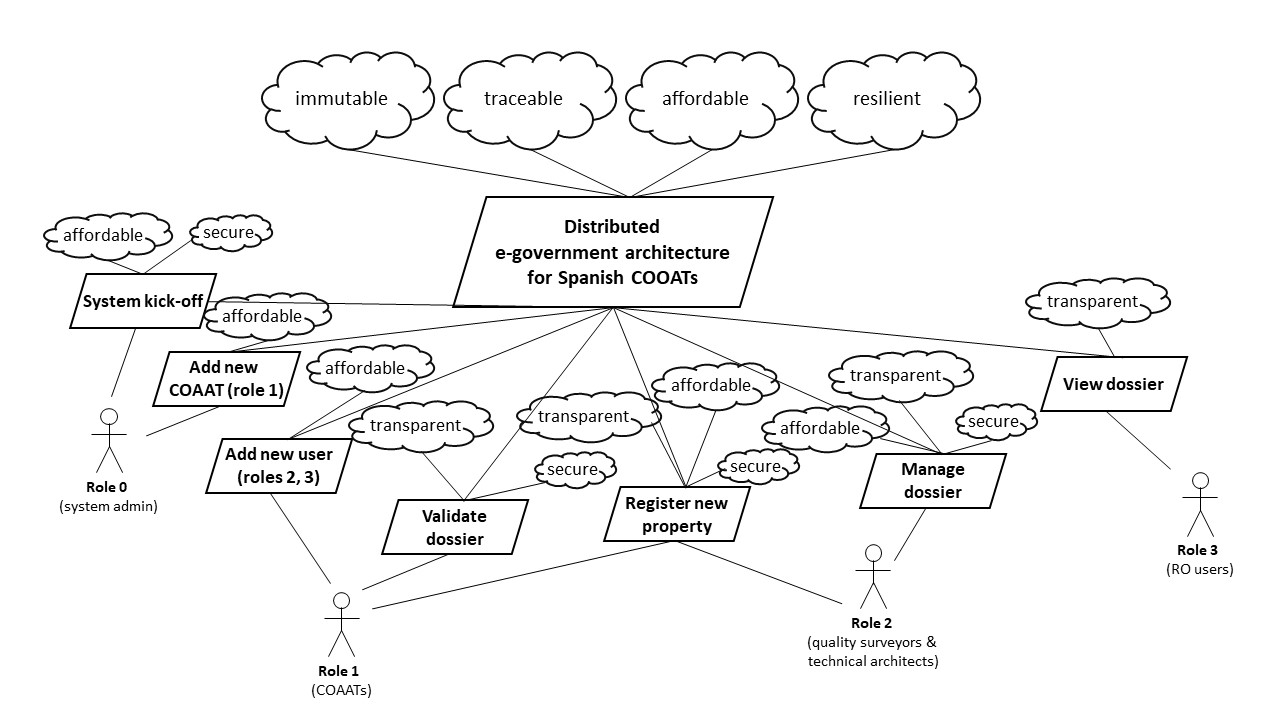}
    \caption{COAATChain system value proposition and its FLG model. admin: administrator; RO users: read-only users}
    \label{fig:COAAT-svp-flg}
\end{figure*}

As outlined in previous sections, the main value proposition of COAATChain (see Figure~\ref{fig:COAAT-svp-flg}) is \textbf{to provide a distributed e-government architecture for Spanish COAATs}. As the platform harnesses the blockchain technology, it inherits its main features, namely, \emph{immutability}, \emph{traceability}, and \emph{resiliency}. Also, \emph{affordability} will be ensured due to the relatively low fees introduced by the use of blockchain platforms (more on this in Section ``Design evaluation''), but also because the proposed solution is engineered to keep the amount of information stored in the blockchain to the minimum, while also guaranteeing the immutability, traceability, and accessibility of the documentation.

From the main value proposition and the tasks every stakeholder must be able to accomplish, the first level goals (FLGs) of the COAATChain platform have been derived in Figure~\ref{fig:COAAT-svp-flg}.
The \emph{System kickoff} goal deploys the system on the selected blockchain platform. 
The \emph{Add new COAAT} goal incorporates new COAATs into the system.  
The \emph{Add new user} goal provides the capability of creating the remaining user roles in the system (on one hand, quality surveyors and technical architects, and on the other hand, read-only users). 
The \emph{Register new property} goal inserts the details of a property into the system. Properties can be associated with one or more dossiers, while only one of them might be open at a certain moment. 
The \emph{Manage dossier} goal enables the inclusion of a new dossier into the system, as well as the addition of documentation associated with it. 
The \emph{Validate dossier} goal is one of the critical goals of the system because only the COAAT stakeholder can validate (or not) the dossier, after examining its associated documentation. 
Lastly, the \emph{View document} goal enables the visualization of the documentation held by any COAAT by generic read-only stakeholders.

The most relevant quality goals have also been identified per every FLG in Figure~\ref{fig:COAAT-svp-flg}. These represent the non-functional system properties associated with each of them. The quality goal \emph{affordable} refers to the feasibility that deployment costs can be assumed by the COAATs. \emph{Secure} means that there will be controls to prevent unauthorized manipulation of certain assets. Finally, the main aim of the system is to enable \emph{transparent} access to documentation for authorized stakeholders.
Moreover, the quality goals associated with the main value proposition (namely, \emph{immutable}, \emph{traceable}, \emph{resilient}, \emph{affordable}, discussed above) are inherited by all the FLGs derived from it. 

The different FLGs have been assigned to their corresponding stakeholder, following the detected actions that each of them is required to perform. For example, Role 3 stakeholders (read-only users) are only associated with the goal \emph{View document} while  Role 1 stakeholders (COAAT administrators) are associated with three goals: \emph{Add new user (Role 2, 3)}, \emph{Validate dossier} and \emph{Register new property}.

\subsection{FLG implementation details}

The \emph{System kickoff} goal consists of the deployment on the blockchain platform of a \textbf{contract factory} that will implement the FLGs \emph{Add new COAAT}, \emph{Add new user} and \emph{Register new property}.

The \emph{Add new COAAT} and \emph{Add new users} goals consist of the identification through a reliable and secure \textit{Know Your Client} (KYC) strategy of the new user  and its registration in the COAATChain by including the wallet's address in the blockchain, the assigned role and the identification data. Please note that Role 0 will act as the KYC agent for Role 1, and in turn, Role 1 will act as KYC agent for their corresponding Role 2 and Role 3.

The \emph{Register new property} goal is implemented as a method on the contract factory that should be executed by Role 1 or Role 2 stakeholders. A transaction is issued to deploy a new smart contract for the property, i.e., the \textbf{property contract}. The cadastral data of the new property must be provided at contract deployment for identification purposes. This smart contract will implement all the FLGs associated with that property, namely, \emph{Manage dossier}, \emph{Validate dossier}, and \emph{View dossier}.
The \emph{Manage dossier} goal implies entering all the data related to the new dossier into the system, as well as the additional documentation required for COAAT validation. This additional documentation is provided in digitally signed PDF files. To implement \emph{immutability} and \emph{security} in the handling of these files, the system generates a Secure Verification Code (SVC) for each of them. This code must be included by the user in the PDF file before signing it digitally. 
These files are digitally signed by a nation-wide certification entity, more precisely, certificates issued by the ``F\'abrica Nacional de Moneda y Timbre, FNMT'' (National Mint and Stamp Office) are used\footnote{The FNMT root certificates can be downloaded from \url{https://www.sede.fnmt.gob.es/descargas/certificados-raiz-de-la-fnmt}}.

\begin{figure}[t]
    \centering
    \includegraphics[width=0.5\textwidth]{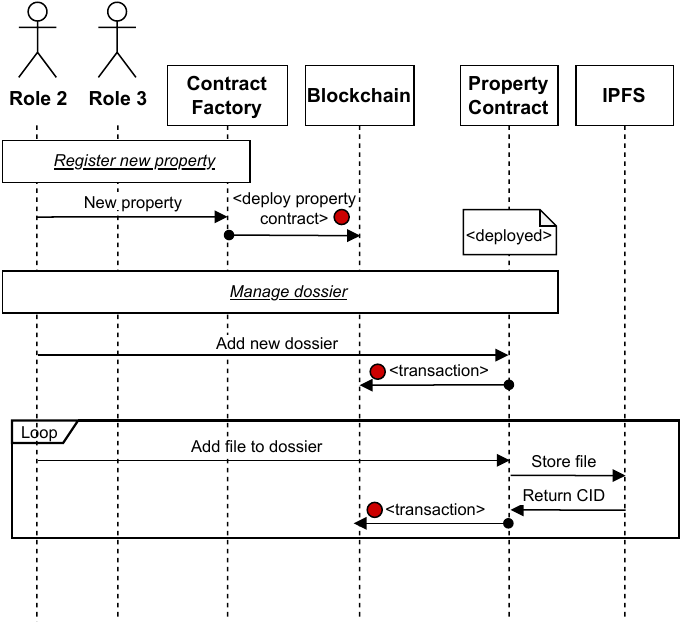}
    \caption{Sequence diagram - Register new property and manage dossier}
    \label{fig:dsecCOAAT}
\end{figure}

As the basis of COAATChain, blockchain enables the creation of a duplicate-free, common database for all COAATs.
However, considering the required store capacity of the PDF files, they are stored off-chain in the IPFS (InterPlanetary File System)~\cite{IPFS,IPFScite}. So, COAATChain just stores in the blockchain the unique CID (Content IDentifier) returned by IPFS  and the SVC of the file. This strategy contributes to achieving the \emph{affordability} quality goal. 
Please note that the contents stored in IPFS (i.e., the PDF files) are public according to Spanish law.
Figure~\ref{fig:dsecCOAAT} illustrates by means of a sequence diagram the interaction among the incumbent entities when performing the actions associated with these roles (Role 1 or Role 2). Transactions stored into the blockchain, which incur an associated cost (fee), are marked with a red circle.

\begin{figure}[t]
    \centering
    \includegraphics[width=0.5\textwidth]{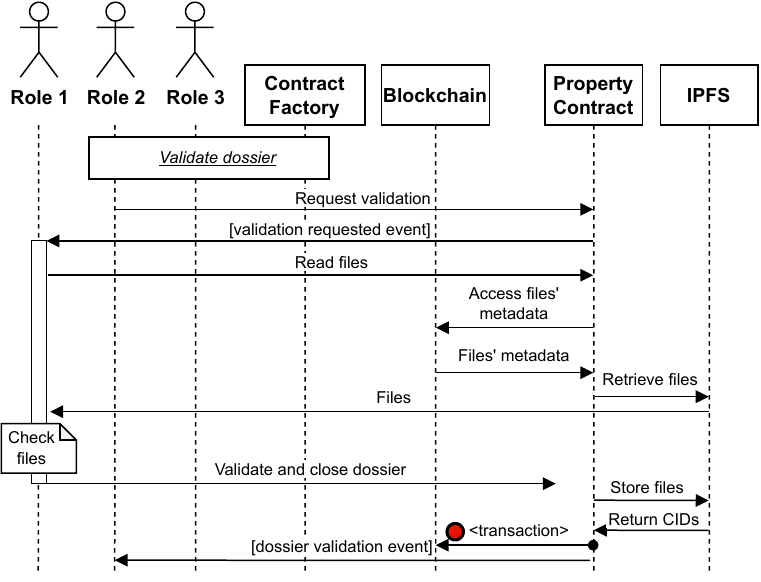}
    \caption{Sequence diagram - Dossier validation}
    \label{fig:dsecCOAAT2}
\end{figure}

The \emph{Validate dossier} goal is triggered when a Role 2 stakeholder requests a dossier validation by calling to the Property contract, as Figure~\ref{fig:dsecCOAAT2} shows. Then, an event is sent to a Role 1 stakeholder, who will access the dossier metadata stored in the blockchain, and retrieve its associated documentation from IPFS using the appropriate CIDs. Once the documentation is processed, the dossier can be either validated or rejected and, new SVCs are included in the PDFs before signing them digitally again. The transaction on the blockchain validating or rejecting the dossier will include the CIDs returned by IPFS when the reviewed PDFs are stored in it. The system implements asynchronous events to alert the stakeholders with Role 1 and 2, respectively, of the inclusion of a dossier in the system and of the dossier status change.

The \emph{View document} function will allow Role 3 stakeholders to access the documentation of a validated dossier by using the SVC code associated with this document by the FLG \emph{Validate dossier}. 
The SVC code is provided to the user through any digital or traditional mechanism (i.e., e-mail or letter).

\section{Design evaluation}\label{sec:cost}

Three of the objectives that were proposed to be achieved with the developed system are traceability, immutability, and resiliency. These features are guaranteed by the blockchain platform itself, where the \emph{System kickoff} goal is performed. This can be verified by accessing the Build and Build (BNB) Blockchain Explorer with the hash of the ``System kickoff" transaction~\cite{kickoffbnc}.

The goal of sharing information between different COAATs when using a public blockchain platform such as the BNB blockchain is intrinsically assured, since all the interested COAATs may take part in COOATChain. Therefore, they will be able to use the smart contract's functions to retrieve data from blockchain and IPFS, and also to enable access to other interested stakeholders (Role 2 and Role 3 users). To avoid duplicity of property records, the smart contract checks whether its cadastral reference exists in the system or not in the ``Register new property" functionality.

Finally, system affordability is proved by comparing the actual fees of the COAAT of Albacete province \cite{coaatfees}, ranging from 12€ ($\approx$13\textdollar) to 655€ ($\approx$705.5\textdollar), with the BNB transaction fees shown in Table \ref{tab:feeBNB} (the hash of each transaction is included). It can be seen as the fees associated with each transaction are fully assumable. They could even be deducted from the current rates charged by the COAATs for the services provided.

\begin{table}[!ht]
\caption{Fee per transaction type.}\label{tab:feeBNB}
\centering
\begin{tabular}{|p{3.1cm} | r@{.}l|r@{.}l|}

\hline
\centering \textbf{Transaction} (Hash) & \multicolumn{2}{c|}{\textbf{BNB}} & \multicolumn{2}{c|}{\textbf{Cost~(\footnotesize \textdollar)}} \\
\hline

Smart contract factory deployment (\href{https://testnet.bscscan.com/tx/0x3a53b5d61dfb01cf3fbf15e91afe0f73c848a76f98af42ddc67d3e2a74a7efde}{0x3a53b..})  & 52 & 50531*$10^{-3}$ & 15 &9 \\\hline
Add COAAT (Role 1) (\href{https://testnet.bscscan.com/tx/0x667beaeeb21d3ec74eb101568a3fd7c94d0d12d4f7842923fb9c38425870579b}{0x667be..}) & 2 &38626*$10^{-3}$& 0&72 \\\hline
Add Quality Surveyor (Role 2) (\href{https://testnet.bscscan.com/tx/0x54500646197a4d8c6c0dc6fcde2b70af4bc91acd5231b9397de51fb8c54879d7}{0x54500..}) & 1 & 77261*$10^{-3}$ & 0&51 \\\hline
Add new property (\href{https://testnet.bscscan.com/tx/0x376256c5bd92c59f630cdb38fb51d7b8dbf7e3dc912c7b4ea396762a0a4b8ce8}{0x37625..})& 30 &27519*$10^{-3}$ & 9&30 \\\hline
Add new dossier (\href{https://testnet.bscscan.com/tx/0x877cfadfde173f0755cacd1d64e7224bd97527897f70efa69ff242159dab8fed}{0x877cf..}) &  4 &09118*$10^{-3}$& 1&18 \\\hline
Add file to dossier (\href{https://testnet.bscscan.com/tx/0xb2fe0c17b4e18a0f1f3a2224ee50d02c609de94a236bebaefb7a3be0f47097f7}{0xb2fe0..}) & 3 &04687*$10^{-3}$& 0&58 \\\hline
Read or check a dossier $^{*}$ & 0& 0 & 0&0 \\\hline
\multicolumn{5}{l}{\scriptsize $^{*}$This action doesn't require a transaction in Binance network.}
\end{tabular}
\end{table}

\section{Conclusions and future work}\label{sec:conclusion}
The proposed COAATChain artifact fulfills the functional and non-functional requirements identified for the problem of providing a suitable e-government architecture for Spanish COAATs. Thanks to the use of blockchain and smart contracts, it provides the desired first level goals (system deployment, user role management, property management, dossier validation, and access to validated documentation), exhibiting features such as immutability, traceability of records, and resiliency of the platform. Moreover, its affordability has also been assessed by deploying it on a real testnet blockchain. 

In the future, it is intended to put the application into production and evaluate its productivity with the reports obtained by the users. 

 To this end, a  proper KYC strategy should be deployed before putting the system into a production state. 
A user-friendly user interface should also be provided to increase adoption by potential users. In addition, support for migrating the service between different blockchain platforms could be offered.

\section{ACKNOWLEDGMENT}

This work was supported under PID2021-123627OB-C52 project, funded by MCIN/AEI/10.13039/501100011033 and by European Regional Development Fund (ERDF), ``A way to make Europe''.

\bibliographystyle{elsarticle-num}
\bibliography{bibliography}

\begin{IEEEbiography}{Ángel F. Alcaide}{\,} 
is currently a part of the software development team with the Official Association of Quantity Surveyors and Technical Architects, Albacete, Spain. His research interests include blockchain technology and e-goverment systems. Alcaide received a computer engineering degree from the University of Castilla-La Mancha. Contact him at \href{mailto://angelfco.alcaide@alu.uclm.es}{angelfco.alcaide@alu.uclm.es}.
\end{IEEEbiography}

\begin{IEEEbiography}{Carlos Núñez-Gómez}{\,} 
is a member of the High-Performance Networks and Architectures Group at the Albacete Research
Institute of Informatics, the University of Castilla-La Mancha, 02071, Albacete, Spain. His research interests include blockchain, information security, and fog computing environments. Nuñez-Gómez received his M.Sc. degree in information and communication security from Universitat Oberta de Catalunya. Contact him at \href{mailto://carlos.nunez@uclm.es}{carlos.nunez@uclm.es}.
\end{IEEEbiography}

\begin{IEEEbiography}{Francisco M. Delicado}{\,} 
is an associate professor with the Department of Computer Engineering, and a member of the High-Performance Networks and Architectures Group at
the Albacete Research Institute of Informatics, the University of Castilla-La Mancha (UCLM), 02071, Albacete, Spain. His research interests include software-defined networking; wireless sensor networks (WSNs); heterogeneous, low-power
WSNs; and cloud networking. Delicado received his Ph.D. degree in computer engineering from UCLM. Contact him at \href{mailto://francisco.delicado@uclm.es}{francisco.delicado@uclm.es}.
\end{IEEEbiography}

\begin{IEEEbiography}{Carmen Carrión}{\,} is an associate professor of computer architecture and technology in the Computing Systems Department, and a member of the High-Performance Networks and Architectures Group at the Albacete Research Institute of Informatics, the University of Castilla-La Mancha, 02071, Albacete, Spain. Her research interests include higher education, resource management schemes, blockchain technology, and quality of service in fog Internet of Things frameworks. Carrion received her Ph.D. degree in physics from the University of Cantabria She is a member of IEEE. Contact her at \href{mailto://carmen.carrion@uclm.es}{carmen.carrion@uclm.es}.
\end{IEEEbiography}

\begin{IEEEbiography}{M. Blanca Caminero}{\,} is an associate professor of computer architecture and technology in the Computing Systems Department, and a member of the High-Performance Networks and Architectures Group at the Albacete Research Institute of Informatics, the University of Castilla-La Mancha (UCLM),
02071, Albacete, Spain. Her research interests include qualityof-service support and efficient resource scheduling in distributed and decentralized systems and technologies (for example, the cloud, fog, edge, blockchain, and so on). Caminero received a Ph.D. degree in computer science from UCLM. She is a member of the IEEE. Contact her at \href{mailto://mariablanca.caminero@uclm.es}{mariablanca.caminero@uclm.es}.
\end{IEEEbiography}

\end{document}